\documentclass[12pt]{iopart}

\usepackage{graphicx} 
\usepackage{ulem}
\usepackage{color}
\begin{document}
\title{Unidirectional Emission in an All-dielectric Nanoantenna}
\author{Tianhua Feng$^1$, Wei Zhang$^1$, Zixian Liang$^2$, Yi Xu$^1$*}
\address{$^1$Department of Electronic Engineering, College of Information Science and Technology, Jinan University, Guangzhou, 510632, China}
\address{$^2$College of Electronic Science and Technology, Shenzhen University, Shenzhen, 518060, China}
\ead{e\_chui@qq.com}
\vspace{10pt}
\begin{indented}
\item[]December 2017
\end{indented}
\begin{abstract}
All-dielectric nanoantenna is a promising alternative of plasmonic optical antennas for engineering light emission because of their low-loss nature in the optical spectrum. Nevertheless, it is still challenging to manipulate directional light emission with subwavelength all-dielectric nanoantennas. Here, we propose and numerically demonstrate that a hollow silicon nanodisk can serve as a versatile antenna for directing and enhancing the emission from either an electric or magnetic dipole emitter. As primarily coupled to both electric and magnetic dipole modes of a nanoantenna, broadband nearly-unidirectional emission can be realized by the interference of two modes, which can be spectrally tuned via the geometric parameters in an easy way. More importantly, the emission directions for the magnetic and electric dipole emitters are opposite by controlling the phase difference between the induced magnetic and electric dipole modes of the antenna. Meanwhile, the Purcell factors can be enhanced more than one order of magnitude and high quantum efficiencies can be maintained at visible spectrum for both kinds of dipole emitters. We further validate these unidirectional emission phenomena are robust to small disorder effect of in-plane dipole orientation and location. Our study provides a simple yet versatile platform that can shape the emission of both magnetic and electric dipole emitters. 
\end{abstract}
%
%
%
%
%
\section{Introduction}
Manipulating light emission plays an important role in the nanophotonics \cite{Book}. As a promising candidate to bridge the gap between free space propagation electromagnetic waves and local excitation and the mismatch caused by the scale difference between the size of emitter and the wavelength of emission, optical antennas can boost and direct light emission at subwavelength scales \cite{OptAnt2005science,OpticalAnt}. Taking advantage of breaking the diffraction limit and the giant enhancement of electromagnetic near-field, plasmonic nanostructures can significantly enhance light emission but they still suffer from significant absorption loss \cite{PRL-Novotny, NP-Bowtie}. Besides of enhancing the light emission, manipulating the direction of the emission beam is also very important and has a lot of applications \cite{OL-Alaee,LPR2016}. Inspired by a typical antenna design in radiofrequency, optical Yagi-Uda antenna has been demonstrated to be capable of directing light emission, which is crucial for improving the collection efficiency of weak optical signal \cite{Yagi-Uda,Yagi-Hofmann,Engheta-PRB2007}. In the conventional configuration of optical Yagi-Uda antenna, the feed element is driven by the optical dipole emitter that is close to it. The so-call ``reflector" and ``director" elements with different sizes are polarized by the feed one and then provide the anti-phase and in-phase far-field interference that facilitates directional light emission, respectively. To provide suitable phase difference, the separation between elements needs to be well controlled and consequently the whole antenna has a footprint that is larger than the working wavelength. It has been proposed and demonstrated that an ultracompact plasmonic nanoantenna can be realized by utilizing the interference effect from the induced dipole modes of nanopatches or nanoparticles \cite{PRL2006-Ultracomp, NL2009-Ultracomp, PRB2010-Ultracomp}. In addition to optical antennas, unidirectional emission has also been demonstrated for whispering-gallery mode under geometric symmetry breaking \cite{LPR2016-Xiao}. Unidirectional emission could also facilitate the free-space coupling of energy, which could be important in many applications, such as nanoscale optical spectroscopy, bright single-photon sources and the recently demonstrated chaos-assisted broadband momentum transformation of light, etc \cite{NatComm2011, RMP-Lodahl, Science-Xiao}.

Despite of the promising enhancement effect for light emission, the low quantum efficiency due to the considerable intrinsic loss of plasmonic nanoantenna still limit their applications in the optical spectrum. Hybrid plasmonic-photonic structures were also proposed to improve the quantum yield \cite{PRL-Xiao, PRA-Xiao, NL-Tong}. Recently, dielectric nanostructures were demonstrated to be better alternatives for realizing low loss optical antennas \cite{OE2012-Rolly, OE-DieMD, PRB2015-Mie,APL-Isabelle, PRAppl-2016, NL2017, Die-Andery}. It has been recognized that dielectric nanoparticles support both magnetic dipole (MD) and electric dipole (ED) resonances that can realize a large amount of fascinating phenomena \cite{Die-Andery, Die-Jacob, NP-isabelle, yang, PRL2017-THFeng, PRL2017-wei}. When the light emitter locates at the vicinity of the dielectric nanoantennas, it can couple to both dipole modes and even high-order modes, leading to the enhancement of emission. Meanwhile, in the far-field region, the radiation pattern can be shaped, which is resulted from the interference effects of induced modes of the nanoantenna. Although most of the nanoantennas mentioned above can exhibit the functionality of directional light emission, an ultracompact all-dielectric nanoantenna with deterministically controlled electric and magnetic dipole modes for enhancing and directing the emission from both the MD and ED emitters is still a challenge \cite{Nanoscale, LPR2015-Krasnok, APL-MQiu}. 

To date, it has been widely investigated for the case of controlling the radiation direction from an ED emitter \cite{2013PRL-Qiu,OL-WSha,ED-Review}. It should be pointed out that the manipulation of ED emission has attracted more research efforts than the MD one as the ED type light sources widely existed in conventional light emitting materials, such as quantum dots and dye molecules, etc. In contrast to the ED emission, MD emission is usually orders of magnitude weaker in most of the natural materials. However, there are still some materials with comparable MD emission, such as rear-earth ions ($Eu^{3+}$ and $Er^{3+}$). Therefore, tailoring the MD emission by nanostructures also received raising research efforts recently \cite{LPR-MDReview}. To significantly enhance the MD emission, plasmonic structures as well as all-dielectric resonators with magnetic resonance have been proposed and demonstrated \cite{OL2011,PRL-Giessen,Diabolo,OL-Meta, JPCC-Dimer, OL2016-Feng, ACSPhoton2017}. Nevertheless, little effort has been paid into the directional MD emission in all-dielectric nanostructures which can still maintain large Purcell factor (PF) and high quantum efficiency(QE) at the same time. 

Here, we propose and numerically demonstrate that a hollow-shaped silicon nanodisk (HSNA) can enhance light emission with high QE for both ED and MD emitters. More importantly, the radiation is nearly unidirectional along a specific direction because of the balance between the excited MD and ED modal amplitudes and the relative phase that can be easily tuned by varying the geometric parameters of the nanoantenna. Such kinds of unidirectional emission can be attributed to the far-field interference between the induced ED and MD modes of the nanoantennas excited by different dipole emitters.

\section{Theoretical models}
\subsection{Spontaneous emission enhancement}
The enhancement of light emission can be characterized by evaluating the spontaneous decay rate of a dipole transition, which can be described with Fermi's Golden Rule\cite{Book}
\begin{eqnarray}
\label{eq:Fermi}
\gamma=\frac{2\pi}{\hbar^2} \sum_{f} \left| \left\langle f \right| \hat{H}_I \left| i \right\rangle \right| ^2 \delta (\omega_i-\omega_f).
\end{eqnarray}
where $\left\langle f \right|$ and $\left| i \right\rangle$ are the final and initial states of a transition, and $\hat{H}_I$ is the interaction Hamiltonian. In the dipole approximation, the interaction Hamiltonian can be expressed as $\hat{H}_I=-\hat{\textbf{p}} \cdot \hat{\textbf{E}} -\hat{\textbf{m}} \cdot \hat{\textbf{B}}-\cdots$, where $\hat{\textbf{p}}$ and $\hat{\textbf{m}}$ are the ED and MD moment operators, while $\hat{\textbf{E}}$ and $\hat{\textbf{B}}$ are the electric and magnetic field operators, respectively. The expression of the interaction Hamiltonian here is general, and it should be emphasized that the field operators might consist of both ED and MD modes as well as higher order modes. If a dipole transition of either ED or MD type could simultaneously couple to both ED and MD modes, the degree of freedom to shape the emission of a dipole emitter would be extended. Later, we will show that a hollow silicon nanoantenna with proper design can support both ED and MD modes at the same resonant wavelength, leading to the enhancement of spontaneous emission with one order of magnitude. The enhancement of spontaneous emission can be characterized with Purcell factor ($PF$), which can be evaluated as the ratio of spontaneous decay rates with ($\gamma$) or without ($\gamma_0$) the structures, namely, $PF=\gamma /\gamma_0$. It should be noted that we can calculate $PF$ in the framework of classical electrodynamics in the regime of weak coupling due to the large radiation loss of all-dielectric nanoantennas \cite{Book}. Therefore, we calculated the radiated power of an oscillating dipole with or without all-dielectric nanoantennas to obtain \textit{PF} \cite{OL2016-Feng}.The spontaneous emission can be separated into two parts, i.e., the radiative part and the nonradiative part due to the materials loss of structures. Therefore, the quantum efficiency ($QE$) can be calculated as the ratio of the far-field radiated power and the total emitted power of an oscillating dipole, that is $QE=P_{rad}/(P_{rad}+P_{abs})$ and $P_{abs}$ is the absorption of structure.

\subsection{Unidirectional emission along opposite directions}
\begin{figure}[!h]
\centering
\includegraphics[width=0.8\columnwidth]{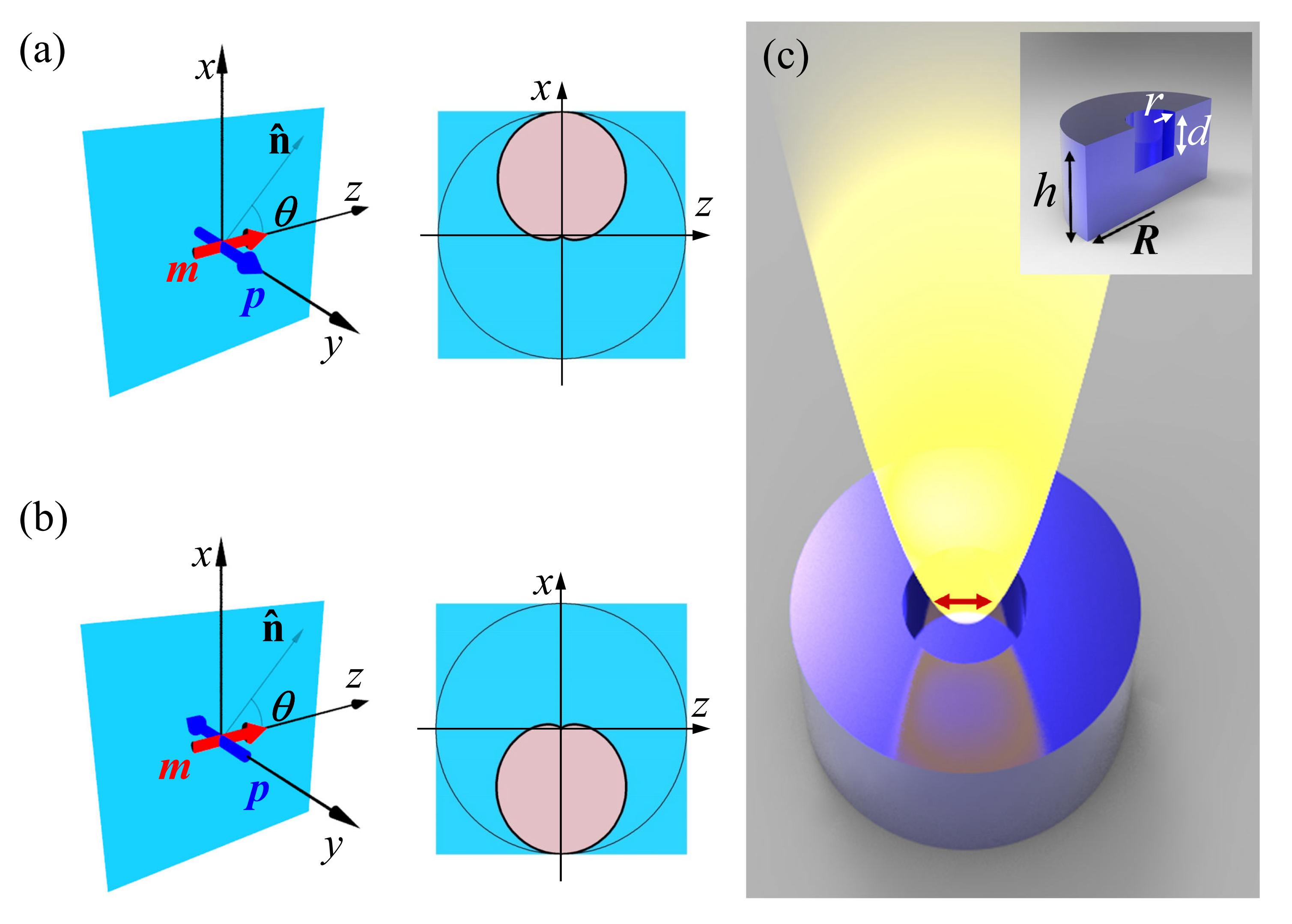}%
\caption{\label{fig:fig1} 
(a) and (b) Left panels: sketch of an electric ($\textbf{p}$) and a magnetic ($\textbf{m}$) dipole arranged in the Cartesian coordinate. $\theta$ is the angle between the spatial unit vector ($\hat{\textbf{n}}$) and the $z$ axis on the $x-z$ plane. Right panels: corresponding 2D far-ﬁeld radiation patterns of power (light pink regions)  when $p=m/c$. The big circle outside is for the reference of isotropy. (c) Sketch of the hollow silicon nanoantenna (HSNA) for unidirectional emission. A dipole emitter (red arrow) is located inside the void so that its far-ﬁeld emission pattern can be shaped, as schematically shown with the yellow region. Inset: Cross-section view of the structure. The height and outer radius of the HSNA are denoted with $h$ and $R$, respectively. The depth and radius of the void are indicated with $d$ and $r$, respectively. }
\end{figure}

When a dipole emitter is located in the vicinity of an antenna, the electromagnetic field from the dipole emitter will polarize the antenna and induce various multipole modes of the nanoantenna with the emitter. If the emitter is very close to the antenna, the induced multipole modes will dominate the overall far-field radiation \cite{PRL-Giessen}. These multipole modes radiate electromagnetic field and their far-field interference can reshape the radiation pattern of a dipole emitter. Therefore, directional emission can be realized by controlling the induced multipolar modes of an antenna. For simplicity, we consider only the ED and MD modes of antenna here and it can be extended to other multipoles in a similar way. Assuming that the induced ED moment $\textbf{p}$ and induced MD moment $\textbf{m}$ of an antenna are along the $+y$ and $+z$ direction respectively, which is schematically showed in figure \ref{fig:fig1}(a), we can express the total electric field in the far-field region on the $x-z$ plane as \cite{Jackson}
\begin{eqnarray}
\label{eq:Etot}
\textbf{E}_{tot}&&=\textbf{E}_p+\textbf{E}_m =\frac{Z_0 c k^2}{4 \pi} \frac{e^{ikr}}{r} \left( p+m/c \sin \theta \right) \hat{\textbf{y}},
\end{eqnarray}
where $Z_0$ is the wave impedance of vacuum, $c$ is the speed of light in vacuum, and $k$ is the  wavenumber. The radiated electric field consists of two parts, one of which is from the induced ED mode and the other is from the induced MD mode. It can be found that when $p=m/c$ is satisfied, unidirectional emission can be realized along the $+x$ direction. The radiated electric field along $-x$ direction $(\theta=-\pi/2)$ vanishes due to the destructive interference between the induced ED and MD modes. A more clear view of the far-field radiation pattern on the $x-z$ plane has been plotted in the right panel of figure \ref{fig:fig1}(a). Similar results can also be obtained on the $x-y$ plane. More interestingly, when one of the two dipole moment changes its direction, the unidirectional emission will turn to the opposite direction. For example, as shown in figure \ref{fig:fig1}(b), the induced ED moment is assumed to be along the $-y$ direction. In this case, the dipolar interference term is modified as $(-p+m/c \sin \theta )$. The far-field radiation becomes zero along the $+x$ direction $(\theta=\pi/2)$ when $p=m/c$, which has also been shown in the right panel of figure \ref{fig:fig1}(b). In order to distinguish these two kinds of unidirectional emission, we describe the ED and MD moments in the former case as \textit{in phase} while in the latter case they are \textit{out of phase} with respect to the defined coordinate. Interestingly, we have found that an ED or MD emitter in the vicinity of a structured silicon nanodisk corresponds to these two kinds of unidirectional emission, respectively.

\section{Results and discussion}
It has been shown that when the induced ED and MD moment of an antenna is in phase with respect to the specific coordinate, the unidirectional emission can be achieved along $+x$ direction while it will be switched to the opposite direction once the two dipole moments are out of phase. Now, we will show how a HSNA can realize such kinds of unidirectional emission for ED and MD emitters. Generally, conventional silicon nanodisks have been shown to support simultaneous ED and MD modes when illuminated with a plane wave. We find that introducing a void in such a nanodisk can bring in more degrees of freedom for controlling its modes properties and then engineering light emission \cite{ACSPhoton2017}. The HSNA is schematically shown in figure \ref{fig:fig1}(c). A cylindrical void is concentrically introduced at the upper part of a nanodisk, as is clarified by the cross-section view in the inset of figure \ref{fig:fig1}(c). The height of the HSNA is $h$ = 86 nm while the outer radius is $R$ = 110 nm. The depth and radius of the void are $d$ = 19 nm and $r$ = 25 nm, respectively. These size parameters are practical and such structures can be fabricated with the state-of-the-art nanotechnology \cite{OE-Polman}. The possible fabrication procedure would include the electron beam lithography and inductively coupled plasma etching. For the light emitters, they might be incorporated via specific solution. For example, quantum dots in the PVA solution \cite{NC-Kivshar}. More importantly, the dipole excitation is physically different from the plane wave case. It provides different degree of freedom to control the excited mode of HSNA. For example, the emitted electromagnetic fields can excite the ED and MD mode of the HSNA in the same spectrum range with comparable strengths, resembling the theoretical conditions derived in Sec.2.2. By locating a ED emitter inside the void, it can be found that its spontaneous emission rate can be enhanced and the corresponding radiation pattern can become unidirectional, as schematically shown in figure \ref{fig:fig1}(c). It should be noted that the symmetry breaking of such structures has also been considered to achieve bianisotropic scattering \cite{PRB-bianiso} and isotropic MD emission \cite{ACSPhoton2017}.

\begin{figure}[]
\centering
\includegraphics[width=0.8\columnwidth]{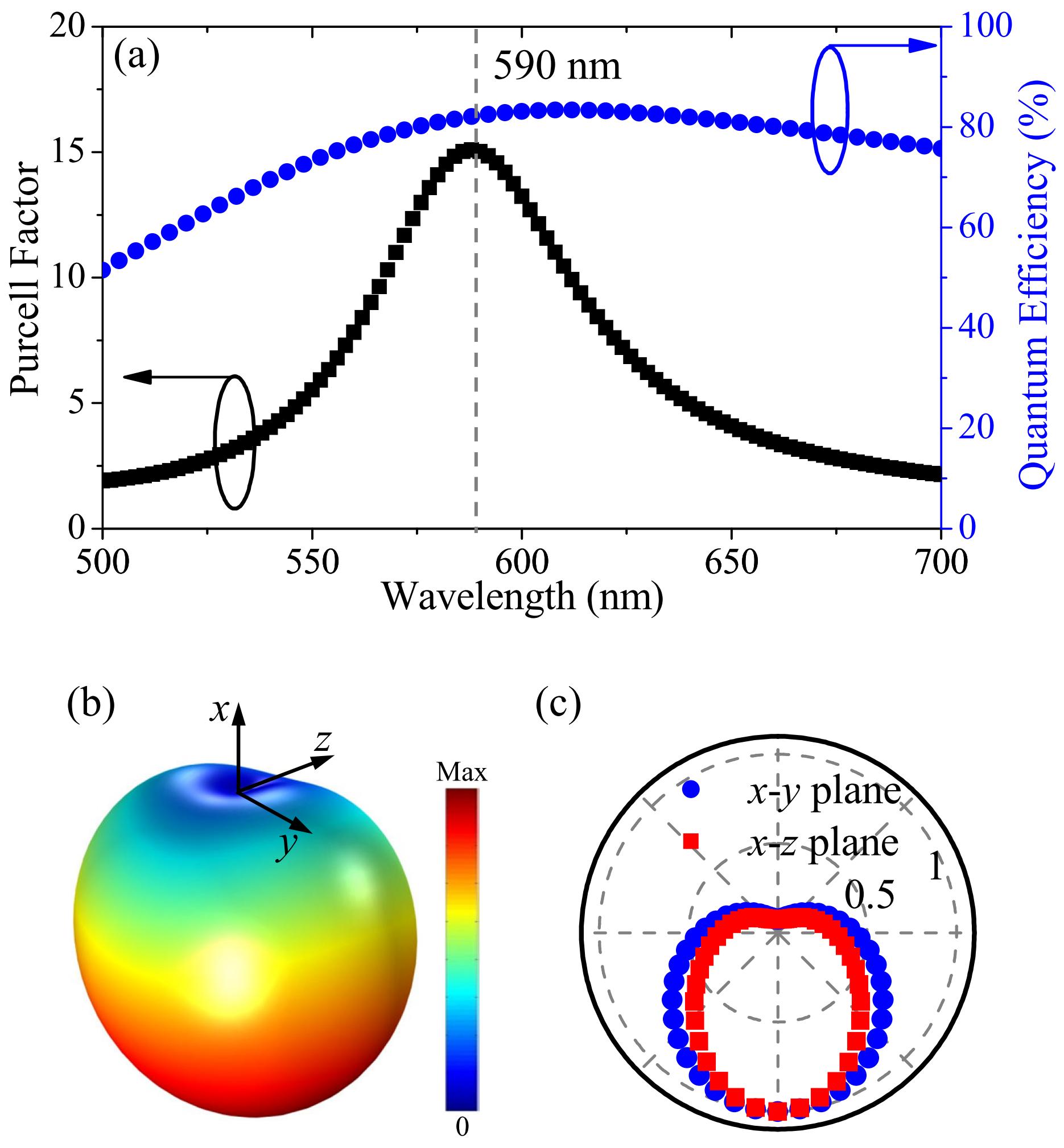}%
\caption{\label{fig:fig2} 
(a) Purcell factor and quantum efficiency for an ED emitter located in the HSNA. The dash line indicates the resonance of 590 nm. (b) 3D far-field radiation pattern at 590 nm. (c) Cross-section views on the $x-y$ and $x-z$ plane of the normalized far-field radiation pattern. }
\end{figure}

The enhancement effect of spontaneous emission rate for an ED emitter that is located inside the void with horizontal dipole moment (along $y$ axis) is shown in figure \ref{fig:fig2}(a). The distance between the ED and the bottom of the void is 10 nm. We can calculate the PF by evaluating the radiated power of the ED that has been mentioned above. The results are obtained from full-wave simulations (Finite Element Method with COMSOL Multiphysics 5.2a). In the simulations, the dielectric constant of silicon is fitted with experimental data \cite{Palik}. We gradually reduce the mesh until converged results are obtained. It can be found from this figure that the PF can increase with one order of magnitude at the resonant wavelength of 590nm. This increased PF can be attributed to the coupling between the ED emitter and both ED and MD modes of the HSNA, which will be demonstrated in the following. The emitted electromagnetic field of the dipole emitter can polarize the HSNA, leading to the excitation of both ED and MD modes of the HSNA. In this way, the dipole emitter can radiate much energy than the case without the HSNA. We can also see that the QE of the ED emission can be as high as 80\% at 590 nm since only dielectric material has been adopted in the HSNA. Importantly, the far-field radiation pattern of the ED is nearly unidirectional along the $-x$ direction, as shown in figure \ref{fig:fig2}(b). We also present its 2D cross-section views on two orthogonal planes, i.e., the $x-y$ and $x-z$ planes, in figure \ref{fig:fig2}(c). This far-field radiation pattern is very similar to the theoretical results that is showed in figure \ref{fig:fig1}(b). As can be seen from this figure, most of the emission energy radiates to the lower half-space. Along the $+x$ direction, the radiation is nearly zero, leading to a large emission ratio between the $-x$ and $+x$ directions. As we will discuss below, it is the far-field interference of the ED and MD modes induced in the HSNA that leads to the unidirectional emission.

\begin{figure}[!t]
\centering
\includegraphics[width=0.8\columnwidth]{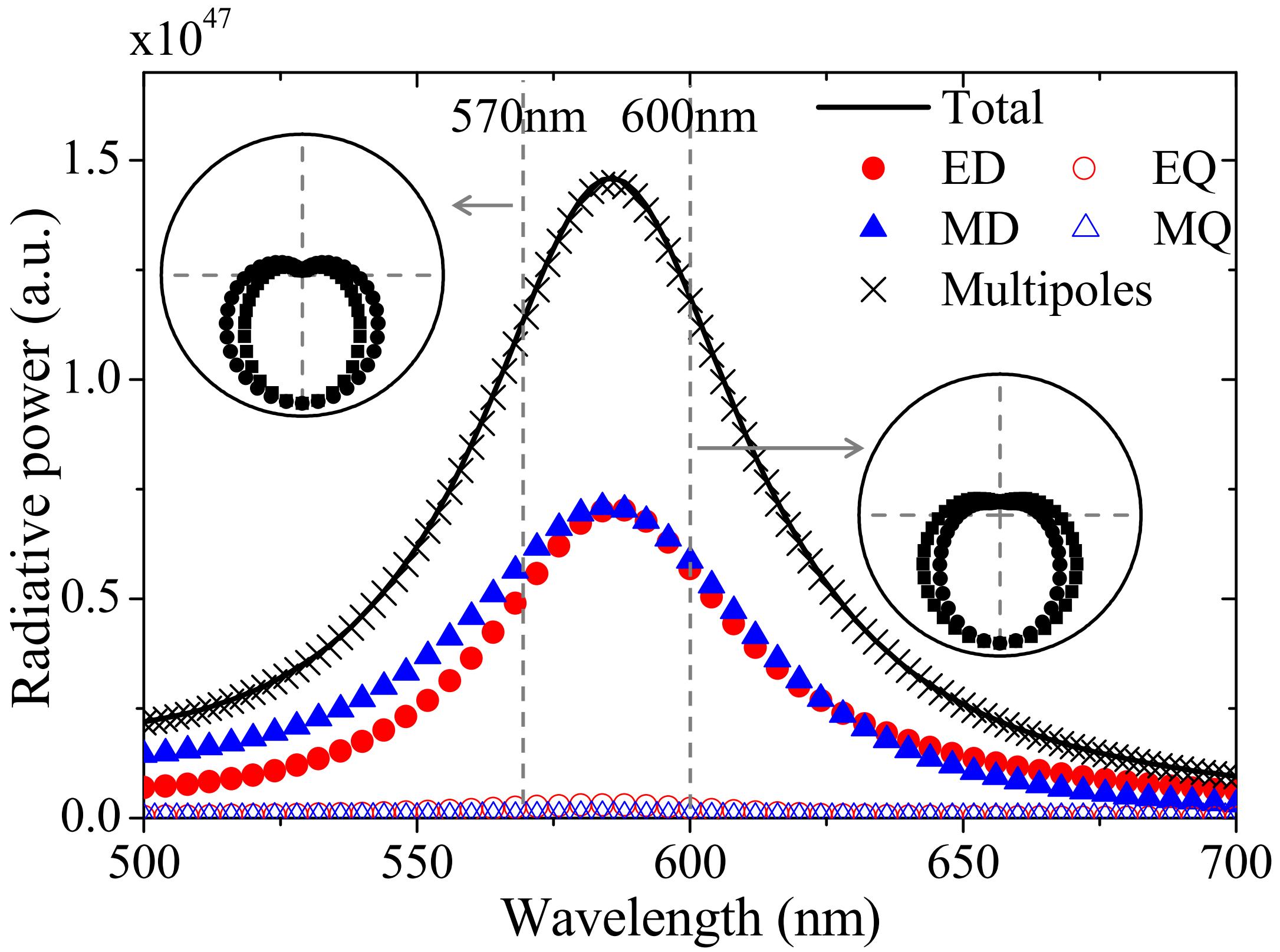}%
\caption{\label{fig:fig3} 
Radiative powers of an ED emitter coupled to the ED, MD, electric quadrupole (EQ) and magnetic quadrupole (MQ) modes of the HSNA. The total radiative power and the sum of radiative powers from multipoles up to quadrupole ones are presented with black solid line and cross symbols, respectively. The insets show the 2D radiation patterns at 570 nm and 610 nm, which are respectively indicated by the gray dashed lines. }
\end{figure}

In order to clarify the mechanism of the directional radiation pattern of the ED emitter, we employ the multipoles expansion of the radiative electromagnetic fields from the ED inside the HSNA (see Appendix A). In this way, we can obtain the coupling between the ED emitter and each radiative multipoles mode by calculating their radiative powers. These results that are up to the quadrupole modes are shown in figure \ref{fig:fig3}. The total radiative power for an ED emitter inside the HSNA, which is integrated in the far-field, is shown with black solid line. It exhibits a resonance at around 590 nm. The contributions of the ED, MD, electric quadrupole (EQ) and magnetic quadrupole (MD) modes to the total emission are also shown in figure \ref{fig:fig3}. As can be seen from these results, the ED and MD modes nearly overlap at around the resonant wavelength. Because of the far-field interference of the excited ED and MD modes, most of the emission power of the ED emitter is directed along the $-x$ direction, leading to unidirectional MD emission along the $-x$ axis. We will demonstrate later that such unidirectional emission can be attributed to the fact that the induced ED and MD moments of the HSNA by the ED emitter are out of phase, as we have theoretically discussed above. In addition, it can be noticed that the balance between the excited ED and MD modes by the ED emitter radiation among the wavelength range 570 nm - 610 nm promise directional emission at this wavelength range. It should be pointed out that the MQ mode of the HSNA is also slightly excited by the ED emitter, which leads to small deviations comparing to the theoretical results that is showed in figure \ref{fig:fig1}. In order to benchmark our numerical results, we sum up the radiative powers of the multipole modes up to quadrupole and the result is shown by cross symbols in figure \ref{fig:fig3}. It is nearly overlaps with the total radiative power integrated from the far field, indicating that the contributions of other higher order multipoles to the total radiative power are negligible. As we are mainly working with dipole modes of the HSNA that supports the relative broadband resonance, the directional emission is more broadband than the one utilizing higher order modes \cite{OL-Alaee}. For instance, we show the 2D far-field radiation patterns at 570 nm and 610 nm in the insets of figure \ref{fig:fig3}, respectively. The directional radiation of the ED emitter can be maintained. 

\begin{figure*}[t]
\centering
\includegraphics[width=1\columnwidth]{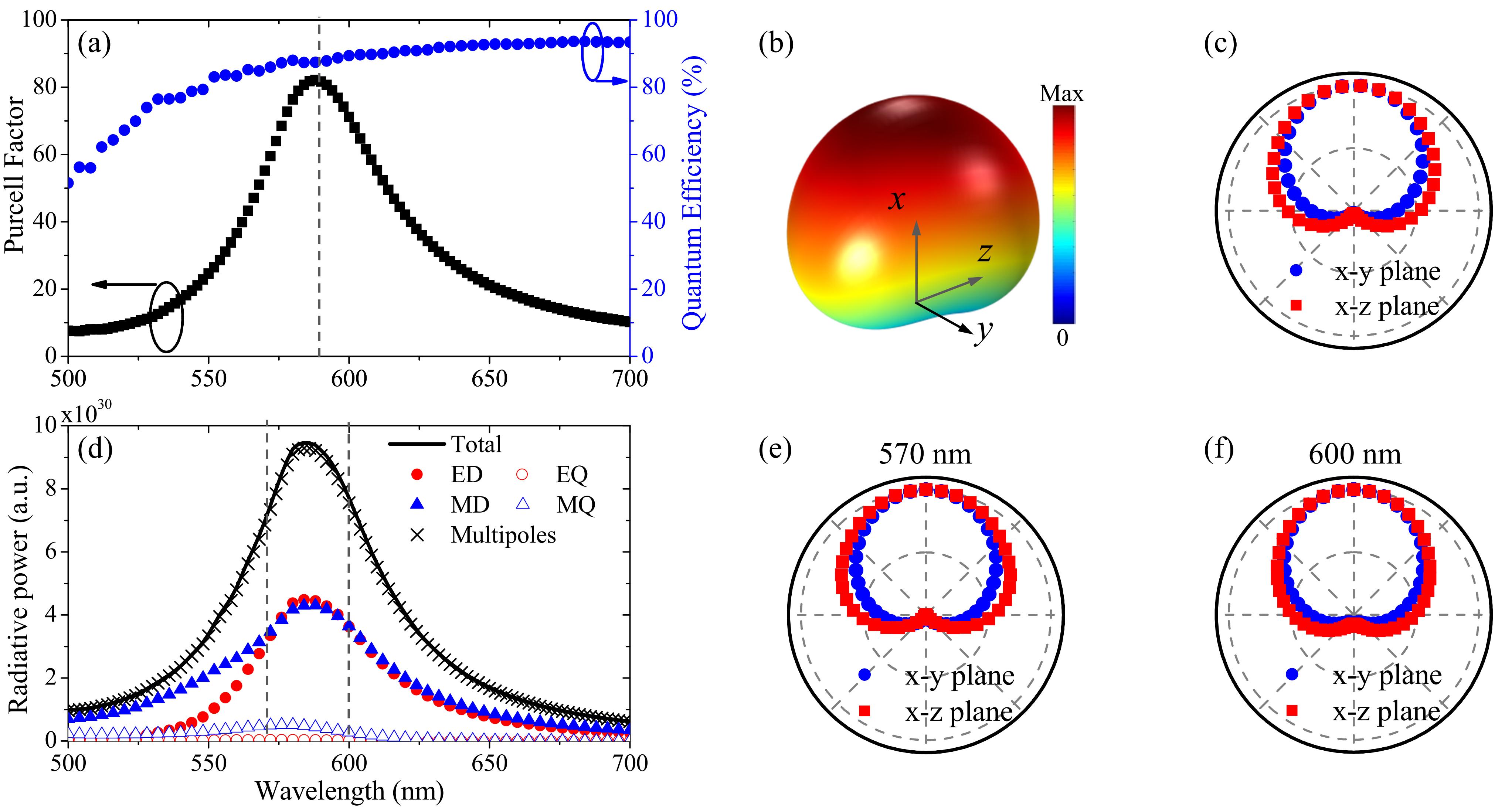}
\caption{\label{fig:fig4} 
(a) Purcell factor and quantum efficiency for a MD emitter located in the HSNA, the geometry parameters are $R$ = 110 nm, $h$ = 80 nm, $d$ = 30 nm and $r$ = 25 nm. The dash line indicates the resonance of 590 nm. (b) 3D far-field radiation pattern at 590 nm. (c) Cross-section views on the $x-y$ plane and $x-z$ plane of the normalized far-field radiation pattern shown in (b). (d) Radiative powers of a MD coupled to four kinds of multipole modes. The sum of radiative powers from these multipoles is also presented with cross symbols. (e) and (f) Cross-section views on the $x-y$ and $x-z$ planes of the normalized far-field radiation pattern at 570 nm and 600 nm, respectively. }
\end{figure*}

More interestingly, this mechanism can also be applied to direct MD emission. In order to excite both ED and MD modes at the same resonant wavelength with MD emitter, we slightly modify the geometric parameters of the HSNA. The height of the HSNA is changed to $h$ = 80 nm while keeping the outer radius $R$ = 110 nm. The depth of the void is modified as $d$ = 30 nm while the radius is still the same, namely, $r$ = 25 nm. We place a MD emitter at the same position with the same dipole orientation in the HSNA. As showed in figure \ref{fig:fig4}(a), the PF can be over 80 and the QE is also larger than 85\% at the resonant wavelength. The far-field radiation patterns at 590 nm are plotted in figure \ref{fig:fig4}(b) and (c). Interestingly, the emission direction of MD is along $+x$ direction, which is opposite to that of ED emission. This can be attributed to the different relative phase of excited ED and MD modes by different types of dipole emitters, which will be explained in more detail later. We also investigate the coupling between the MD emitter and the modes of the HSNA to interpret the unidirectional emission. It can be found in figure \ref{fig:fig4}(d) that the emission is also mainly resulted from the interference between ED and MD modes. At around the resonant wavelength, the radiative powers for both ED and MD modes are nearly the same, leading to the realization of unidirectional emission for MD emitter. Meanwhile, the broadband properties of unidirectional MD emission can still be maintained, indicating by the far-field radiation patterns at 570 nm and 600 nm that are shown in figure \ref{fig:fig4}(e) and (f), respectively. 

\begin{figure}[]
\centering
\includegraphics[width=0.8\columnwidth]{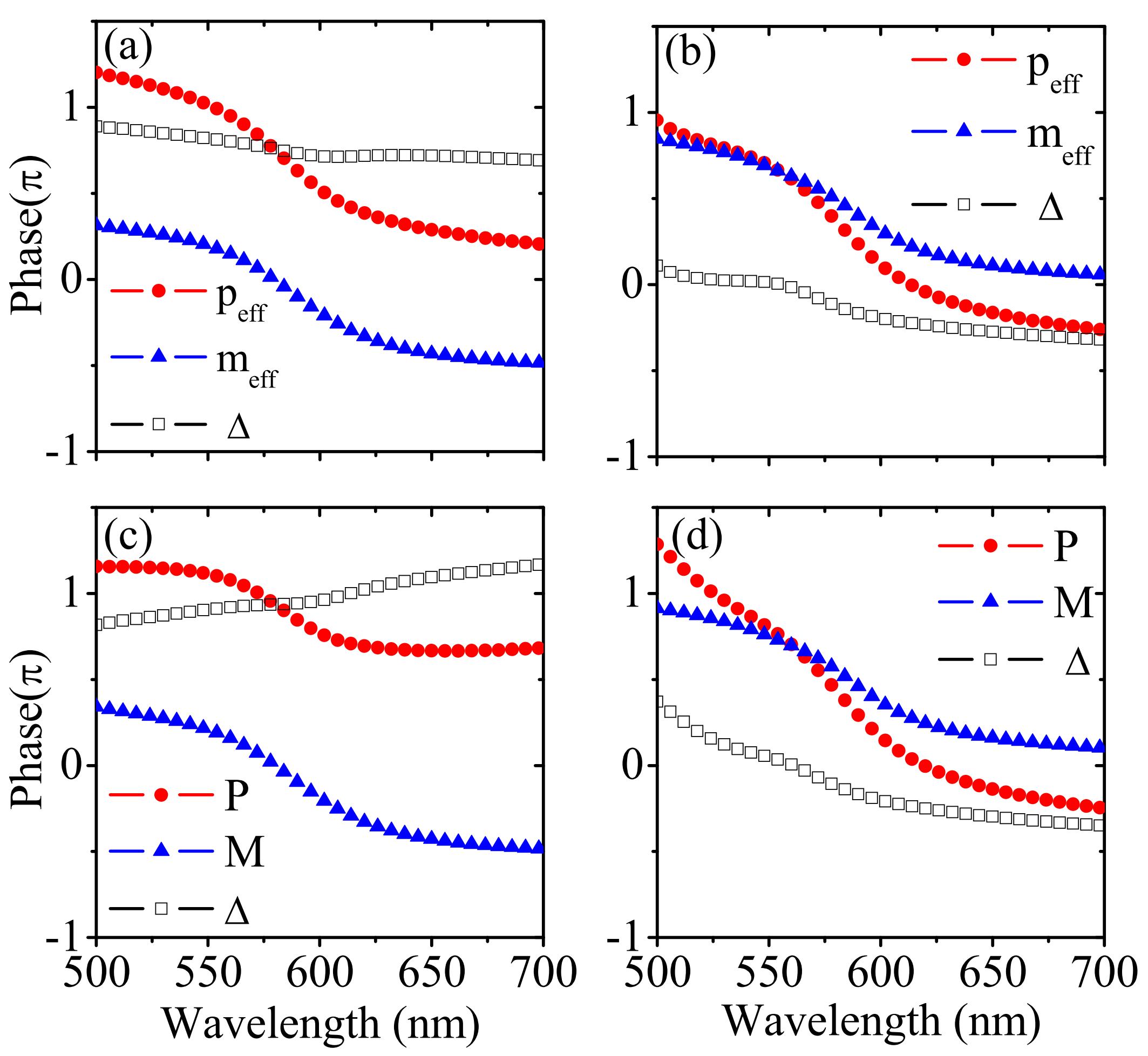}%
\caption{\label{fig:fig5} 
(a) and (b) Phases of effective ED ($\textbf{p}_{eff}$) and MD ($\textbf{m}_{eff}$) moments for an ED and a MD emitters in the HSNA, respectively. These results are extracted from the fields outside the HSNA. (c) and (d) Phases of Cartesian ED ($\textbf{P}$) and MD ($\textbf{M}$) that are extracted from the induced current in the HSNA in the cases of ED and MD emitters excitations, respectively. The phase differences ($\Delta$) for all cases are also presented. }
\end{figure}

It is necessary to explore the origin of opposite directional emission for the ED and MD emitters. As we showed in the theoretical model, different phases of ED and MD with respect to the specific coordinate will result in different emission direction. The unidirectional emission will be along $+x$ direction when the ED and MD modes are in phase while it would be along $-x$ direction once the ED and MD modes are out of phase. With the condition of $p=m/c$ is fulfilled, the direction of dipole emission mainly depends on the phase differences between the ED and MD modes excited in the HSNA. To explicitly extract the phase relationship between the excited ED and MD modes in the HSNA, we derive their effective dipole moment from the multipoles expansion coefficients, which is described in Appendix B. The results are shown in figure \ref{fig:fig5}(a) and (b) for ED and MD emitter, receptively. We can see that for the case of an ED emitter, there are a phase difference that are closed to $\pi$ for the effective ED ($\textbf{p}_{eff}$) and MD ($\textbf{m}_{eff}$) moment at around 590 nm. This phase difference indicates that the out-of-phase ED and MD modes lead to the directional emission along the $-x$ direction. In contrast, for the case of MD emitter, the phase difference between both effective dipole moments is small, indicating that the ED and MD modes are in phase. Therefore, the emission direction of MD emitter is along $+x$ direction. To confirm our conclusion, we also present the corresponding results of Cartesian multipole expansion based on the induced currents of the HSNA (see Appendix C) other than utilizing the radiative fields outside. The case of ED emitter is shown in figure \ref{fig:fig5}(c) while that of MD emitter is showed in figure \ref{fig:fig5}(d). It can be found that the results are similar and we can come to the same conclusion that the excited dipole modes are out of phase for ED emitter while they are in phase for MD emitter. 

\begin{figure}[!ht]
\centering
\includegraphics[width=0.8\columnwidth]{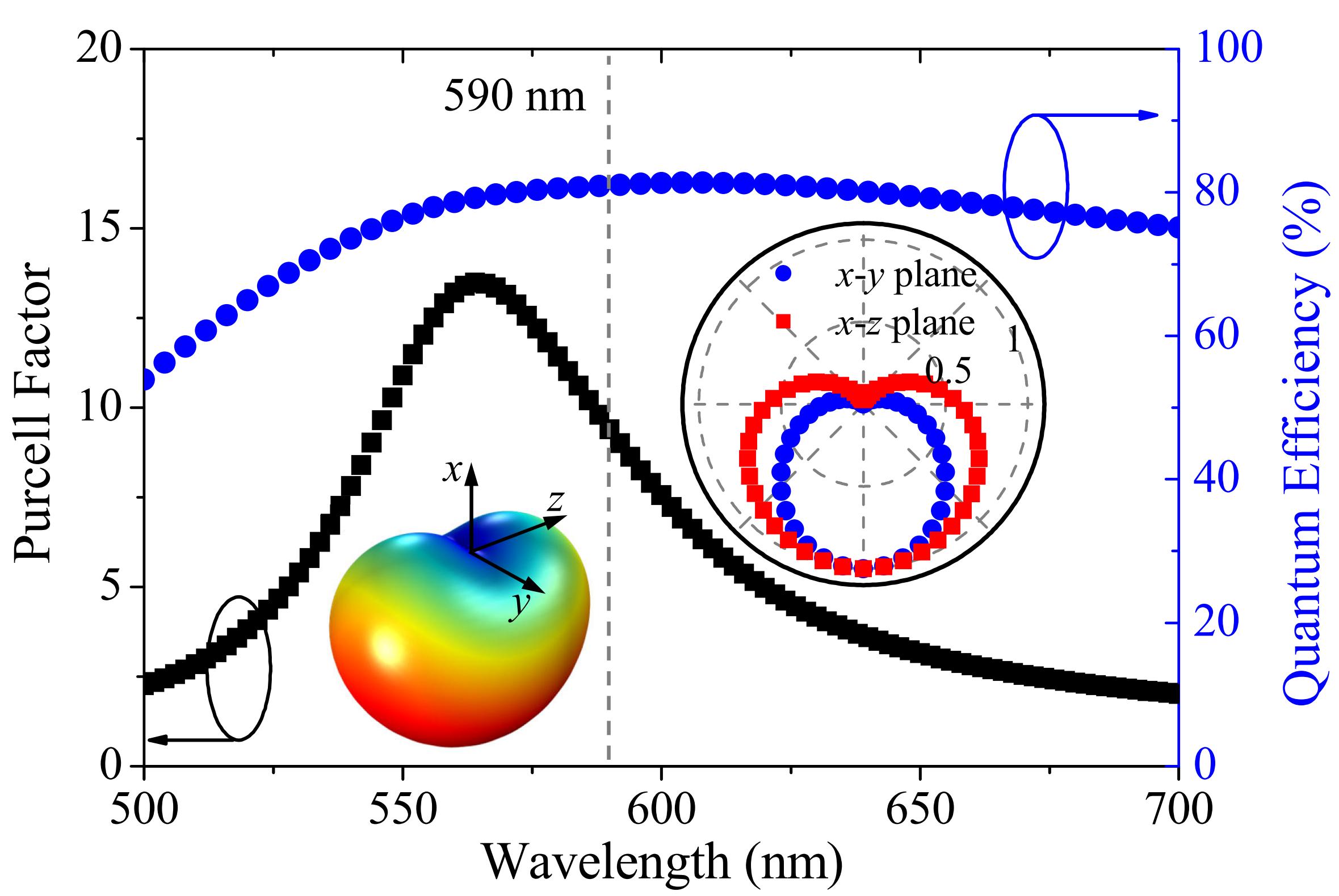}
\caption{\label{fig:fig6} 
Purcell factor and quantum efficiency for an ED emitter placed at the same position in the HSNA designed for the unidirectional MD emission. The inset shows the far-field radiation pattern at 590 nm for both 3D and 2D cases. }
\end{figure}

Particularly, the ED and MD emission may mix in the same emitters and the MD branching ratio is usually low \cite{2012PRB-Zia}. It is essential to study how an individual HSNA modulates both ED and MD emission simultaneously. The question is whether an individual HSNA can separate the emissions of ED and MD in opposite directions. Clarifying this question would help to distinguish the ED and MD transitions via collecting light along opposite directions. As a simple example, we study the case of the latter HSNA that is designed for unidirectional MD emission. A horizontal ED emitter is placed at the same position as that for the MD in the HSNA. As can be found in figure \ref{fig:fig6}, the resonance of the HSNA excited by the ED emitter is blue shifted. The PF at 590 nm is smaller than 10 and the QE is about 80\%. We have presented both 3D and 2D far-field radiation pattern in the insets, which indicates nearly unidirectional emission for ED emitter can be achieved in the opposite direction in contrast with MD emitter shown in figure \ref{fig:fig4}. Consequently, the HSNA can largely promote MD emission along one direction while separate the ED emission along the opposite direction, providing the possibility to increase the ratio of MD emission to ED emission.

\begin{figure}[!b]
\centering
\includegraphics[width=1\columnwidth]{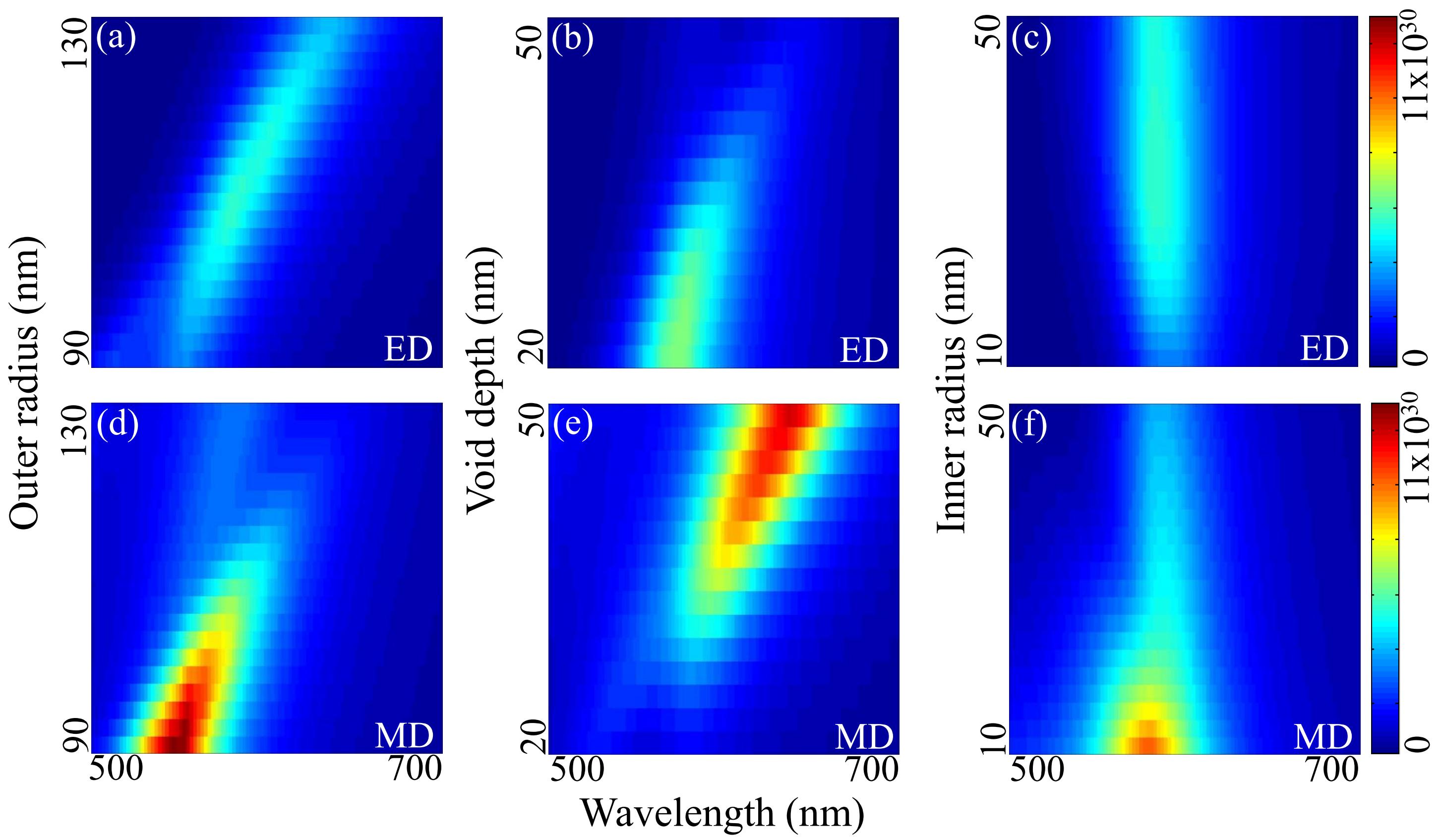}
\caption{\label{fig:fig7} 
Radiative powers coupled to the ED (top panels) and MD (bottom panels) modes of the HSNA vs. structural parameters. (a) and (d): varying the outer radius of the HSNA; (b) and (e): varying the depth of the void; (c) and (f): varying the radius of the void. All plots share the same color scale. }
\end{figure}

In the following, we will study how the structure parameters affect both ED and MD modes, providing practical information for experimental demonstrations. Three key geometric parameters of the HSNA have been studied, i.e., the outer radius, the depth and the radius of the air void. As an example, we focus on the case of MD emission and the emitter is kept with horizontal dipole moment and 10 nm away from the bottom of the void. The results are shown in figure \ref{fig:fig7}. It can be found that both ED and MD modes will red shift as increasing the outer radius of the HSNA, as shown in figure \ref{fig:fig7}(a) and (d). Therefore, the wavelength for directional emission can be tuned by varying the outer radius of the HSNA. Meanwhile, the coupling strength for MD drops faster than that for ED. We then only increase the depth of the void while fixing the bottom thickness of the HSNA, and the results are showed in figure \ref{fig:fig7}(b) and (e). In this case, the radiative powers for both ED and MD modes red shift but the former one drops while the latter one increases. Consequently, the resonant wavelength and the coupling strength for ED and MD can mainly be tuned by modifying the outer radius and the depth of the void. We also present the results for varying the radius of the void in figure \ref{fig:fig7}(c) and (f). The resonant wavelength of both ED and MD modes is almost the same. The coupling strength for ED mode changes a little but that for MD modes drops faster as the inner radius increases from 10 nm to 20 nm.

\begin{figure*}[]
\centering
\includegraphics[width=1\columnwidth]{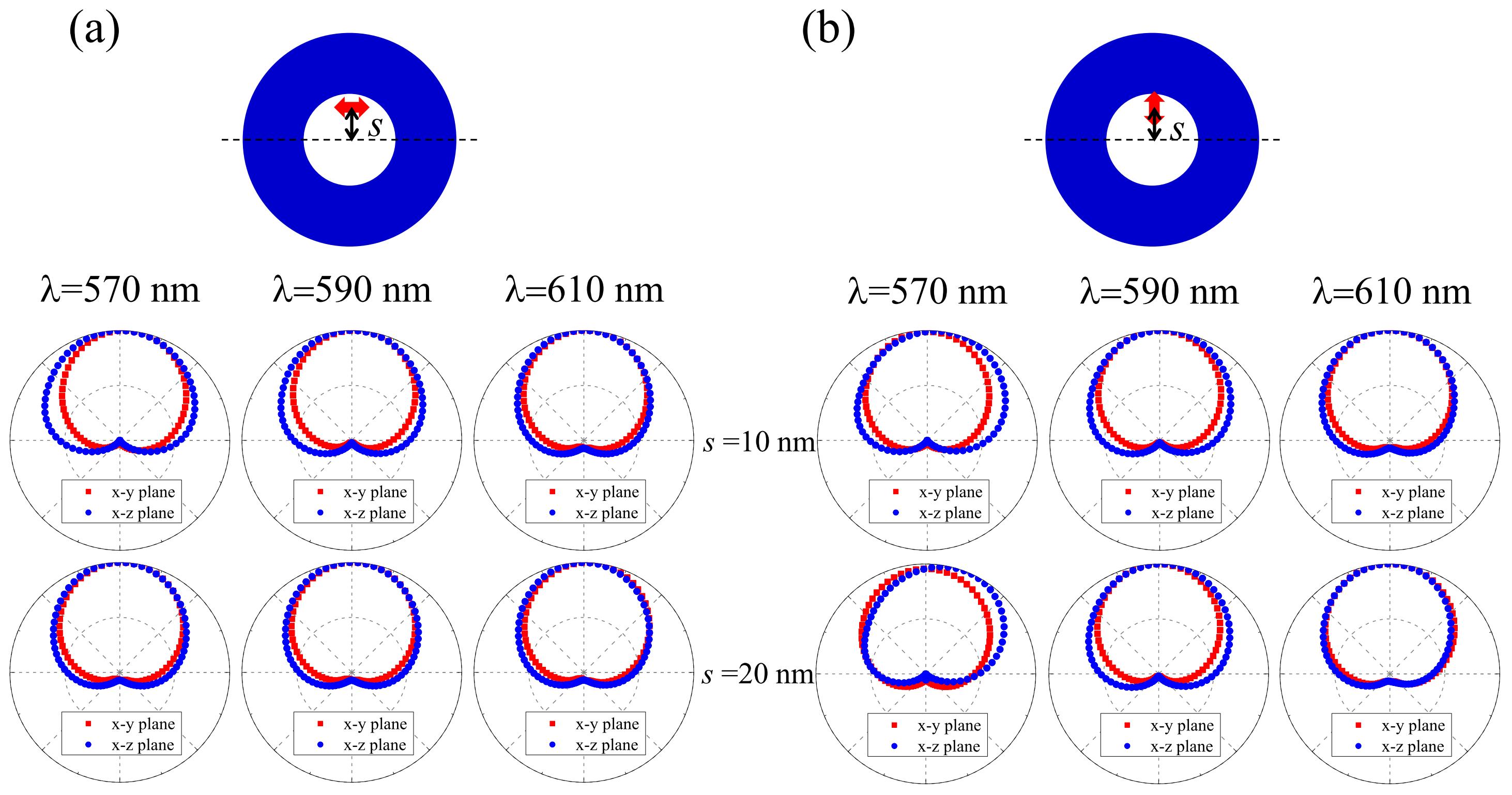}
\caption{\label{fig:fig8} 
(a) 2D normalized far-field radiation patterns for a MD emitter deviating from its original position along the perpendicular direction with respect to the dipole moment with different distances $s$. (b) The case for the MD emitter deviating along the direction of the dipole moment. Three spectral positions are considered for both cases. The schematics are shown on top of the figures. }
\end{figure*}

We also discuss the tolerance of the directional radiation effect by examining the dependence of the directional MD emission on the location of emitter in the void of the HSNA. Without loss of generality, we consider two cases for a MD located on the horizontal plane that is 10 nm away from the void bottom. In the first case, the MD emitter deviates the center of the HSNA with a distance $s$ along the direction perpendicular to its dipole moment, as schematically shown in figure \ref{fig:fig8}(a). We have presented the cross-section views of the radiation patterns for the situations $s$ = 10 nm and $s$ = 20 nm, respectively. At three typical wavelengths, namely, 570 nm, 590 nm and 610 nm, the directivity of radiation patterns are all preserved. In the second case, we deviate the MD emitter from the center of the HSNA along its dipole orientation. Similarly, the directivity of the MD emission can also be maintained in a relative broadband range, as shown in figure \ref{fig:fig8}(b). Therefore, the directional radiation of the MD emitter in the HSNA is robust, which is of significant importance for experimental demonstrations. It should be pointed out that the wavelength range supporting directional emission of MD emitter in the HSNA is sufficient large to cover the emission spectrum of $Eu^{3+}$, which is also useful for practical application.

\section{Conclusion}
We propose and numerically demonstrate that the unidirectional dipole emission for both ED and MD emitters can be achieved utilizing an all-dielectric nanoantenna. The key point is to balance the amplitudes and the phases of ED and MD modes in the HSNA excited by the dipole emitter. Meanwhile, the PF can be maintained over 80 with a high quantum efficiency of 85\% for the MD emission. In particular, the nanoantenna can also distinguish the ED and MD emissions by directing their emissions in opposite directions. The emitter-position dependence of the directional radiation is also studied and it is shown that the directional emission is quite robust for small in-plane displacement of the MD emitter. Such kinds of directional radiation from dipole emitter could improve the collection efficiency of light emission and facilitate the unambiguous study of light-matter interaction. 

\section*{Acknowledgements}
This work was supported by the National Natural Science Foundation of China (Grant Nos. 11704156, 11674130, 91750110 and 11574216); Natural Science Foundation of Guangdong Province, China (Grant Nos. 2016A030306016 and 2016TQ03X981); the Leading Talents of Guangdong Province Program (No. 00201502); the Fundamental Research Funds for the Central Universities (No. 21617346). The authors would like to thank A. E. Miroshnichenko, Wei E. I. Sha and C. W. Qiu for useful discussions. 

\section*{Appendix A: Multipoles expansion of radiated electromagnetic fields}
For a dipole emitter in the vicinity of an antenna, the emission can be coupled to various modes of the antenna. In order to clarify the coupling strength between the dipole emitter and each multipoles mode, it is necessary to extract the multipoles coefficients from the radiated electromagnetic field outside the antenna as \cite{Jackson}
\begin{eqnarray}
\label{eq:eq3}
a_E(l,m)=-\frac{k}{Z_0 h_l^{(1)}(kr)\sqrt{l(l+1)}}\int Y^*_{lm} \textbf{r}\cdot\textbf{E}\textit{d}\Omega,
\end{eqnarray}
\begin{eqnarray}
\label{eq:eq4}
a_M(l,m)=\frac{k}{h_l^{(1)}(kr)\sqrt{l(l+1)}}\int Y^*_{lm} \textbf{r}\cdot\textbf{H}\textit{d}\Omega,
\end{eqnarray}
where $a_E$ and $a_M$ are the electric and magnetic multipoles coefficients, $ k $ is the wavenumber, $ h_{l}^{(1)}(x) $ is the Hankel function of the first kind, $ Y_{lm} $ is the scalar spherical harmonic, and $ Z_{0} $ is the impedance of vacuum. The integral runs over the total solid angle. With these multipoles coefficients, we can obtain the coupling strength between the dipole emitter and each radiative multipoles mode by calculating their corresponding radiative power as
\begin{eqnarray}
\label{eq:eq5}
P_{E,M}(l,m)=\frac{Z_0}{2k^2}|a_{E,M}(l,m)|^2.
\end{eqnarray}
\section*{Appendix B: Effective dipole moment of the ED and MD modes}
As discussed in Appendix A, the radiative electromagnetic fields can be expanded into different multipoles. However, it is not straightforward to find out the phase information of each mode from the coefficients extracted with multipole expansion method. Explicitly, we here derive the effective dipole moments of both ED and MD modes with the multipole coefficients. In general, the radiative electric field of the ED mode can be expressed as \cite{Jackson}
\begin{eqnarray}
\label{eq:eq6}
\textbf{E}=Z_0\sum_{l,m}\left[\frac{i}{k}a_E(l,m)\nabla \times h^{(1)}_1(kr) \textbf{X}_{lm} \right],
\end{eqnarray}
where $\textbf{X}_{lm}$ is the normalized spherical harmonic. For the proposed HSNA, we have found that $a_E$ for $m=\pm 1$ are much smaller than that for $m=0$ although it is lack of spherical symmetry. Therefore, we here only consider $a_E(1,0)$ of the ED mode for the sake of simplicity. In the far-field region, the electric field can be written as:
\begin{eqnarray}
\label{eq:eq7}
\textbf{E}=a_{ED} \frac{\sqrt{3}}{2\sqrt{2 \pi}} \frac{-iZ_0}{k} \frac{e^{ikr}}{r} \sin \theta \hat{\textbf{e}}_\theta,
\end{eqnarray}
where $a_{ED}=a_E(1,0)$ and $\hat{\textbf{e}}_\theta$ is the unit vector of the polar angle with respect to the dipole moment. Suppose that this electrical far-field was generated by an effective ED with its analytical expression
\begin{eqnarray}
\label{eq:eq8}
\textbf{E}_{d}=p_{eff} \frac{-k^2}{4 \pi \epsilon_0} \frac{e^{ikr}}{r} \sin \theta \hat{\textbf{e}}_\theta,
\end{eqnarray}
we can then obtain the effective dipole moment of the ED mode as
\begin{eqnarray}
\label{eq:eq9}
\textbf{p}_{eff}=\frac{\sqrt{6\pi} i }{c k^3}a_{ED}\hat{\textbf{e}},
\end{eqnarray}
where $\hat{\textbf{e}}$ is the unit vector of the dipole moment. Similar derivation can be made for the effective dipole moment of the MD mode by examining its magnetic far-field, which is expressed as
\begin{eqnarray}
\label{eq:eq10}
\textbf{m}_{eff}=\frac{-\sqrt{6\pi} i}{k^3}a_{MD}\hat{\textbf{e}}.
\end{eqnarray}
Consequently, we can directly obtain the phase information of the effective dipole moment of both ED and MD modes induced in the nanoantennas.
\section*{Appendix C: Cartesian multipole expansion}
Other than extracting the effective dipole moments from the radiative electromagnetic fields, Cartesian multipole expansion based on the induced currents can also provide the related information of induced multipole moments\cite{Zheludev-Science}, serving as an alternative for verifying the phase relationship between the ED and MD modes of the proposed HSNA. The Cartesian ED and MD moments can be calculated as:
\begin{eqnarray}
\label{eq:eq11}
\textbf{P}=\frac{1}{-i \omega} \int \textbf{J}(\textbf{r}) d^3r,
\end{eqnarray}
\begin{eqnarray}
\label{eq:eq11}
\textbf{M}=\frac{1}{2c} \int \textbf{r} \times \textbf{J}(\textbf{r})d^3r,
\end{eqnarray}
where $\textbf{J}=-i \omega \epsilon_0 (\epsilon_r-1) \textbf{E}$ is the induced currents in the structure and $\textbf{r}$ is the position vector with the origin at the center of the HSNA here.

\end{document}